\documentclass[12pt,preprint]{aastex}

\usepackage{graphicx, color}
\usepackage{dcolumn}
\usepackage{bm}
\usepackage{booktabs}
\usepackage{natbib}
\usepackage{lscape}

\newcommand {\vect}[1]{\mbox{\boldmath $#1$}}

\newcommand {\inty}[2]{\int_{#1}^{#2}}

\newcommand {\dif}[3][]{\frac{d^{#1}#2}{d#3^{#1}}}
\newcommand {\pdif}[3][]{\frac{\partial^{#1}#2}{\partial#3^{#1}}}

\newcommand {\lsim}{\hspace{0.3em}\raisebox{0.4ex}{$<$}\hspace{-0.75em}\raisebox{-.7ex}{$\sim$}\hspace{0.3em}}
\newcommand {\gsim}{\hspace{0.3em}\raisebox{0.4ex}{$>$}\hspace{-0.75em}\raisebox{-.7ex}{$\sim$}\hspace{0.3em}}

\makeatletter
\def\mart{\@ifnextchar[{\mart@@}{\mart@}}
\def\mart@@[#1]#2{\sqrt[#1]{\mathstrut{#2}}}
\def\mart@#1{\sqrt{\mathstrut{#1}}}
\makeatother
\newcommand {\Alfven}{Alfv\'{e}n}
\newcommand{\myemail}{minoshim@jamstec.go.jp}

\newcommand{\RHESSI}{\it RHESSI}
\newcommand{\TRACE}{\it TRACE}

\newcommand{\TPP}{trap-plus-precipitation}

\long\def\symbolfootnote[#1]#2{\begingroup%
\def\thefootnote{\fnsymbol{footnote}}\footnote[#1]{#2}\endgroup}

\begin{document}

\title{Coronal Electron Distribution in Solar Flares: Drift-Kinetic Model}
\shorttitle{Flare Electron Distribution}
\author{Takashi Minoshima\altaffilmark{1}, Satoshi Masuda\altaffilmark{2}, Yoshizumi Miyoshi\altaffilmark{2}, and Kanya Kusano\altaffilmark{1,2}}
\altaffiltext{1}{Institute for Research on Earth Evolution, Japan Agency for Marine-Earth Science and Technology, 3173-25, Syowa-machi, Kanazawaku, Yokohama 236-0001, Japan}
\altaffiltext{2}{Solar-Terrestrial Environment Laboratory, Nagoya University,
Furo-cho, Chikusa-ku, Nagoya 464-8601, Japan}
\shortauthors{Minoshima et al.}
\email{\myemail}

\begin{abstract}
Using a model of particle acceleration and transport in solar flares, we investigate the height distribution of coronal electrons by focusing on the energy-dependent pitch-angle scattering.
When pitch-angle scattering is not included, the peak heights of loop-top electrons are constant, regardless of their energy, owing to the continuous acceleration and compression of the electrons via shrinkage of magnetic loops. 
On the other hand, under pitch-angle scattering, the electron heights are energy dependent; intermediate energy electrons are at a higher altitude, whereas lower and higher energy electrons are at lower altitudes. 
{This implies that the intermediate energy electrons are inhibited to follow the shrinking field lines to lower altitudes because pitch-angle scattering causes efficient precipitation of these electrons into the footpoint and their subsequent loss from the loop.
}
This result is qualitatively consistent with the position of the above-the-loop-top hard X-ray (HXR) source that is located above coronal HXR loops emitted by lower energy electrons and microwaves emitted by higher energy electrons.
Quantitative agreement with observations might be achieved by considering primary acceleration before the onset of loop shrinkage and additional pitch-angle scattering via wave-particle interactions.
\end{abstract}


\section{Introduction}\label{sec:introduction}
{{
One of the most important features of solar flares is the above-the-loop-top 20-50 keV hard X-ray (HXR) source located higher than coronal $\lsim 20$ keV HXR loops, discovered by \cite{1994Natur.371..495M}. 
Although this discovery is regarded as compelling evidence of magnetic reconnection, the origin of the above-the-loop-top source itself remains a debatable issue in solar flare physics.
Following this discovery, the height distribution of coronal HXR sources in multiple energy bands has been observed and studied using the {\it Reuven Ramaty High Energy Solar Spectroscopic Imager} ({\RHESSI}). \cite{2003ApJ...596L.251S}, \cite{2004ApJ...612..546S}, and \cite{2008ApJ...676..704L} discovered two distinct sources near and away from the solar surface. The energy of the source near the surface increases toward higher altitudes. This energy-dependent height distribution is similar to that of the above-the-loop-top source. The energy of the source away from the surface increases toward lower altitudes. Such observations have been regarded as evidence of the formation of a current sheet between these two sources. Assuming symmetry with respect to the central current sheet, the height distribution of the source away from the surface is also similar to that of the above-the-loop-top source.
}}

Numerous studies on the above-the-loop-top HXR source have been carried out thus far.
\cite{1996ApJ...464..985A} analyzed the time-of-flight differences of HXR light curves, and they suggested that the HXR source itself is the electron acceleration site. 
Plasma turbulence, which is potentially generated by a fast flow associated with magnetic reconnection, has been discussed as a possible particle accelerator \citep{1996ApJ...461..445M,2004ApJ...610..550P}.
\cite{1998ApJ...495L..67T} suggested that the HXR source is the region where electrons are accelerated by fast magnetosonic shocks produced by the collision of a fast flow with underlying magnetic loops. 
On the other hand, \cite{1999ApJ...522.1108M} analyzed HXR spectra at both the corona and footpoint, and they concluded that the HXR sources are a consequence of the magnetic trap of parent electrons at the loop top and the subsequent precipitation into the footpoint via pitch-angle scattering (termed as {\TPP} by \cite{1976MNRAS.176...15M}).
The {\TPP} model has been numerically studied by \cite{1998ApJ...505..418F}, \cite{2004A&A...419.1159K}, and \cite{2008ApJ...673..598M}, exhibiting significant confinement of electrons at the loop top (but not necessarily above the loop).


However, previously proposed models have not successfully determined why the HXR source at a particular energy band is seen {\it only above} the loop and not in the loop, as in the case of other energy bands.
Recently, \citet[hereafter M2010]{2010ApJ...714..332M} developed a comprehensive model for particle acceleration and transport in solar flares on the basis of the drift-kinetic theory. 
This model can describe the time evolution of the particle distribution function associated with the evolution of flaring electromagnetic fields. 
As compared to previous models, this model rigorously treats particle acceleration, transport, and dissipation processes in a realistic time-varying flare environment.

To understand the nature of the above-the-loop-top HXR source, we investigate the distribution of coronal electrons by using the model of M2010.
In particular, we focus on the effect of the energy-dependent pitch-angle scattering due to Coulomb collisions on the resulting electron height distribution. 
In Section \ref{sec:model}, we briefly describe our model. 
Simulation results are presented in Section \ref{sec:results}. 
We summarize the paper and discuss the nature of the above-the-loop-top source in Section \ref{sec:summary-discussion}.

\section{Models}\label{sec:model}
The model solves the drift-kinetic Fokker-Planck equation,
{\normalsize
\begin{eqnarray}
\pdif{f}{t}+\nabla \cdot \left(\dif{\vect r}{t}f\right) + \pdif{}{\mu}\left(\dif{\mu}{t} f\right) + \pdif{}{\gamma}\left(\dif{\gamma}{t} f\right) = \pdif{}{\mu}\left( D_{\mu \mu} \pdif{f}{\mu}\right),\label{eq:2}
\end{eqnarray}
}where $f(\vect{r},\gamma,\mu)$ is the electron distribution function, $\vect{r}$ is the position in configuration space, and $(\gamma,\mu)$ are the Lorentz factor and pitch-angle cosine, respectively. 
The term $d\vect{r} /dt$ consists of the free stream along the magnetic field line and the electric field drift, and the terms $d\gamma /dt$ and $d\mu /dt$ consist of the betatron, inertia drift, and magnetic mirror forces (see Equations (7)-(9) in M2010). 

In addition, we include the pitch-angle scattering due to Coulomb collisions as a diffusion term on the right-hand side.
We use the pitch-angle diffusion coefficient given by the Rosenbluth-MacDonald-Judd theory \citep{1957PhRv..107....1R},
{\normalsize
\begin{eqnarray}
D_{\mu \mu} = \frac{K n (1-\mu^2)}{2 \beta^3} \left\{1+\left(1-\frac{1}{2 x^2}\right){\rm erf}(x) + \frac{1}{\mart{\pi} x} \exp\left(-x^2\right) \right\}, \label{eq:3}
\end{eqnarray}
}where $K \simeq 7.48 \times 10^{-13}$, $\beta = \left(1-\gamma^{-2}\right)^{1/2}$, $x=\left\{\left(\gamma-1\right)/T\right\}^{1/2}$, and ${\rm erf}(x)$ is the error function.
The first term (unity) within the brackets corresponds to the collision with ambient cold ions, and the remaining terms correspond to the collisions with electrons themselves.
In Equation (\ref{eq:3}), it is assumed that the target electrons are almost isotropic in velocity space.
This is a good approximation because the target electrons are relatively slow and collisional.  
$n$ and $T$ are the electron number density and temperature (average energy) in units of the rest mass energy, respectively. They are calculated from the distribution function at each time step,
\begin{eqnarray}
n\left(\vect{r}\right) &=& \inty{-1}{1} d\mu \inty{1}{\infty}d\gamma f,\\
T\left(\vect{r}\right) &=& \frac{2}{3n} \inty{-1}{1} d\mu \inty{1}{\infty}d\gamma \left(\gamma-1\right)f.
\end{eqnarray}

We use the same model as that of electromagnetic fields in M2010. 
The magnetic field model was originally proposed by \cite{1995SoPh..159..275L}; it involves the superposition of two-dimensional potential and horizontal fields to impose an X-type neutral line. 
The simulation domain is limited to the area below the neutral line.
On the basis of observations \citep{2002ApJ...565.1335Q}, we change the magnetic field configuration with time to induce electric fields. For further details of the model, see M2010.

In our model, we consider two initial number densities, $10^{10} \; {\rm cm^{-3}}$ (typical value in active regions) and $10^{11} \; {\rm cm^{-3}}$, in order to evaluate the effect of the number density that is used for the calculation of the pitch-angle diffusion coefficient.
The diffusion term is solved by the second-order conservative form of the interpolated differential operator scheme \cite[IDO-CF;][]{2008JCoPh.227.2263I} via implicit time integration.
Other simulation parameters, the initial and boundary conditions, and the numerical method are the same as those in M2010.

\section{Results and Interpretation}\label{sec:results}
\begin{figure}[p]
\centering
\epsscale{.7}
\plotone{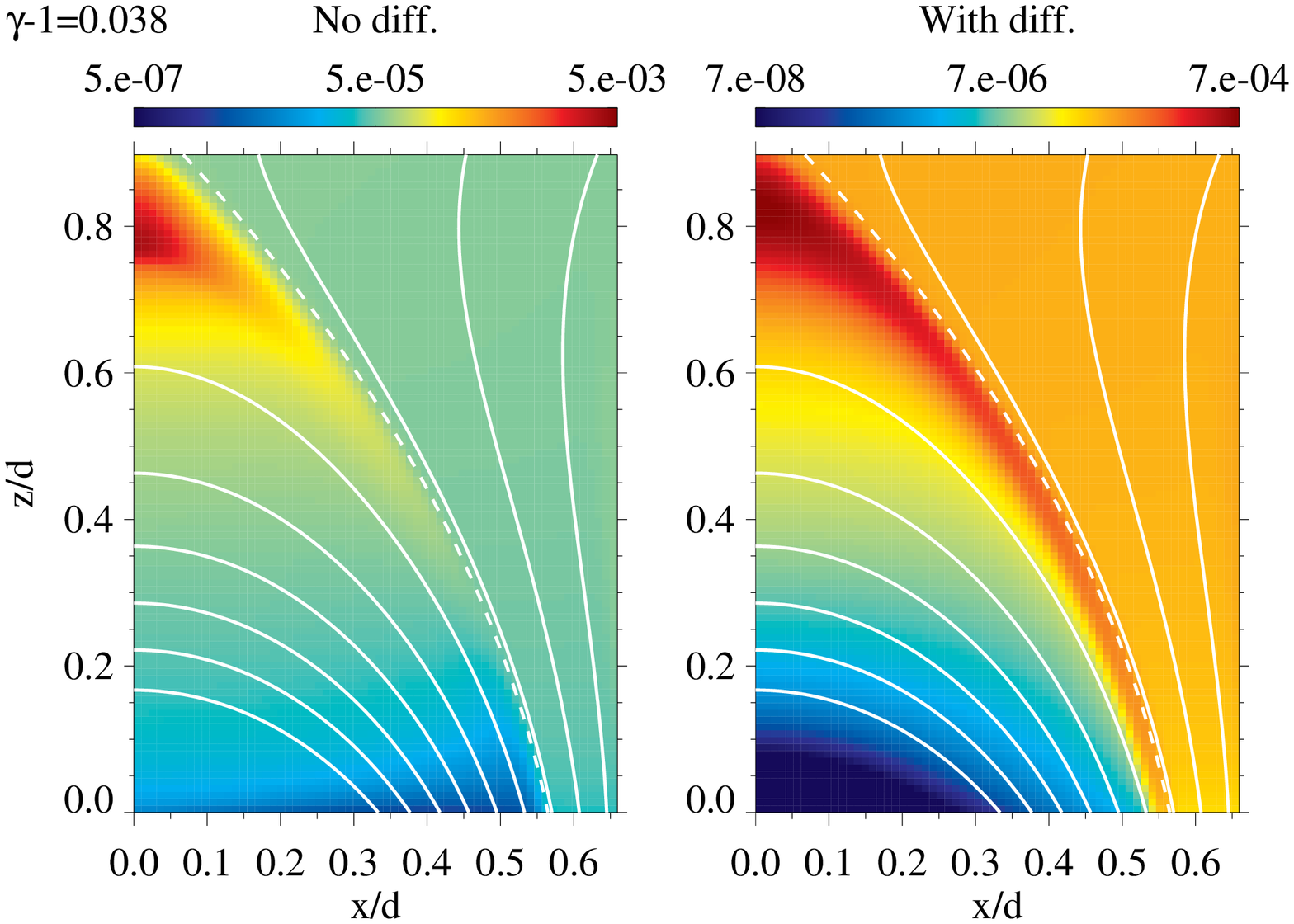}
\caption{Spatial distribution of the number of 20 keV omnidirectional electrons when maximum electric fields are generated (5 s after the flare onset, see M2010 for details). The left and right images show the results without and with pitch-angle diffusion, respectively. The initial number density is $10^{10} \; {\rm cm^{-3}}$. The white lines are the magnetic field lines, and the dashed lines are the magnetic separatrix. The spatial coordinates are normalized by $d = 15 \; {\rm Mm}$.}
\label{fig:fig1}
\end{figure}

Figure \ref{fig:fig1} shows the spatial distribution of the number of 20 keV omnidirectional electrons in the cases without (left) and with (right) pitch-angle diffusion (hereafter, the no-diffusion and diffusion cases, respectively). The initial number density is $10^{10} \; {\rm cm^{-3}}$. In both the cases, the 20 keV electron number is increased around the loop top by betatron acceleration and confinement due to shrinkage of the loops, as discussed in M2010.
In the diffusion case, the electrons are distributed to a greater extent along the field lines toward the lower area, as compared to the no-diffusion case. 
The velocity distribution in the diffusion case (not shown) implies that the loss cone is somewhat filled with electrons (depending on energy), as compared to that in the no-diffusion case (clear loss-cone distribution, Figure 4 in M2010).
The electron number decreases in the bottom area because electrons in the loss cone precipitate into the lower boundary.
In the diffusion case, the number further decreases because electrons tend to enter the loss cone via pitch-angle scattering.

One can see that the peak heights of 20 keV electrons at $x=0$ (loop top) differ in both the cases. The peak height in the diffusion case is higher than that in the no-diffusion case. Figure \ref{fig:fig2} clearly shows the electron height distributions at the loop top. The top and bottom plots show the results with initial number densities of $10^{10}$ and $10^{11} \; {\rm cm^{-3}}$, respectively. In the no-diffusion case (dashed lines), the electron peak heights are constant $(z=0.77)$, regardless of their energy. This can be understood as follows. Any energy electrons are perfectly trapped by the magnetic mirror force without pitch-angle scattering; hence, they are continuously accelerated and compressed via loop shrinkage toward a lower altitude. Therefore, they peak at the same low altitude. 
 
On the other hand, in the diffusion case (solid lines), we find that the electron peak height depends on the energy of the electrons. Lower energy electrons peak at a lower altitude ($z < 0.8$, close to the no-diffusion case), intermediate energy electrons peak at a higher altitude ($z > 0.8$, far from the no-diffusion case), and higher energy electrons peak at a lower altitude ($z < 0.8$, close to the no-diffusion case). The difference in the peak heights is around 1 Mm. The energy of electrons located at a high altitude is higher when the initial number density is larger. 

We interpret the simulation results in terms of a combination of loop shrinkage and the {\TPP} model \citep{1976MNRAS.176...15M}. Unlike the no-diffusion case, electrons can escape from the magnetic trap via pitch-angle scattering. Some electrons that are scattered into a small pitch angle can reach deep into the loop and precipitate into the footpoint; finally, they are lost from the loop. The precipitation (loss) rate can be evaluated by a combination of the bounce frequency in the loop, $\nu_{b} \simeq v/L$, and the pitch-angle diffusion frequency, $\nu_{D} \sim D_{\mu \mu}(\mu=0)$. Here, $v$ is the electron velocity and $L$ is the loop half length. 
When the diffusion frequency is considerably higher/lower than the bounce frequency, the precipitation rate $\nu_{p}$ is approximated by
\begin{eqnarray}
\nu_{p} \simeq 
\left\{
\begin{array} {ll}
\nu_{s} = \left(1-\mu_{c}\right) \nu_{b}, & \;\;\;{\rm for} \;\;\; \left(\nu_{D} \gg \left(1-\mu_{c}\right) \nu_{b} \right), \\
\nu_{w} = \nu_{D}, & \;\;\;{\rm for} \;\;\; \left(\nu_{D} \ll \left(1-\mu_{c}\right) \nu_{b} \right),
\end{array}
\right.
\end{eqnarray}
where $\mu_{c} = \mart{1-B_{LT}/B_{FP}}$ is the loss-cone angle cosine measured at the loop top, and $B_{LT}$ and $B_{FP}$ are the magnetic field strengths at the loop top and footpoint, respectively. These two opposite limiting cases are known as the strong and weak diffusion limits \citep{1969RvGSP...7..379K}.
Figure \ref{fig:fig3} shows the precipitation rate in the strong (dashed line) and weak (dash-dotted line) diffusion limits, and their inverse root mean square (solid line),
\begin{eqnarray}
\nu_{p} = \frac{1}{\mart{\left(1/\nu_{s}\right)^2 + \left(1/\nu_{w}\right)^2}} \label{eq:6},
\end{eqnarray}
where $n=10^{10} \; {\rm cm^{-3}}$, $L=15 \; {\rm Mm}$, and $B_{FP}/B_{LT} = 11$ are the parameters of the loop at $(x,z)=(0,0.8)$.

In the simulation, we consider pitch-angle diffusion via Coulomb collisions. The diffusion frequency is considerably higher than the bounce frequency for lower energy electrons; hence, their precipitation rate is close to the strong diffusion limit $\nu_{s}$. The strong diffusion limit is determined by the electron energy and loop configuration (i.e., the mirror ratio and loop length). In this case, the precipitation rate is lower than the time scale of loop shrinkage (dotted line in Figure \ref{fig:fig3}). The electron loss is low because the precipitation rate is small, and the resulting height distribution is similar to that in the no-diffusion case (green lines in Figure \ref{fig:fig2}).
 Lower energy electrons can follow the shrinking loop to lower altitudes.

For higher energy electrons, the situation is opposite to that of the lower energy electrons. The diffusion frequency is considerably lower than the bounce frequency; hence, precipitation rate is close to the weak diffusion limit $\nu_{w}$. The precipitation rate is also lower than the time scale of loop shrinkage. 
In addition, betatron acceleration due to the shrinkage increases the number of energetic electrons at the loop top. This contribution is more prominent for higher energy electrons because the rate of increase in the electron number $(\partial \ln f / \partial t)$ is proportional to the slope of the energy distribution, which is steeper at higher energy (for Maxwellian, $\partial \ln f / \partial t \propto -\partial \ln f / \partial \gamma = (2K-T)/2KT$, where $K$ is the kinetic energy). Thus, the electron loss is low, and the resulting height distribution is similar to that in the no-diffusion case (pink and red lines in the top plot of Figure \ref{fig:fig2}).
 Higher energy electrons can also follow the shrinking loop to lower altitudes.

As shown in Figure \ref{fig:fig3}, the most efficient precipitation takes place in the case of intermediate energy at which the diffusion frequency is comparable to the bounce frequency (a few collisions during one loop transit). When the precipitation rate is higher than the time scale of loop shrinkage and the rate of increase in the electron number via acceleration, intermediate energy electrons are lost from the loop in the course of the shrinkage. 
{This prevents them from following the shrinking loop to lower altitudes. As a result, the intermediate energy electrons peak at a higher altitude (blue and purple lines in the top and bottom plots of Figure \ref{fig:fig2}).}

When the number density is larger than $10^{10} \; {\rm cm^{-3}}$, the point of intersection of the dashed and dash-dotted lines in Figure \ref{fig:fig3} shifts rightward. 
Thus, the most efficient precipitation takes place at higher energy, and this is consistent with the bottom plot of Figure \ref{fig:fig2}, which shows that the energy of electrons located at a high altitude is higher (20-50 keV, pink and red lines) when the number density is larger ($10^{11} \; {\rm cm^{-3}}$). 

\section{Summary and Discussion}\label{sec:summary-discussion}
We investigated the height distribution of coronal electrons in solar flares by using the drift-kinetic model developed by \cite{2010ApJ...714..332M}. We found that the peak height of loop-top electrons depends on their energy under energy-dependent pitch-angle scattering. Intermediate energy electrons peak at a higher altitude than higher energy and lower energy electrons because of their efficient loss by precipitation into the footpoint via pitch-angle scattering in the course of loop shrinkage. 
{{This prevents them from following the shrinking field lines to lower altitudes.
}}

{{
Observations of coronal HXR sources have indicated that in the HXR energy range, higher energy electrons are located at higher altitudes \citep{1994Natur.371..495M,2003ApJ...596L.251S,2004ApJ...612..546S,2008ApJ...676..704L}. Using microwave images taken by the Nobeyama Radioheliograph (NoRH), \cite{2010AGUFMSH11B1641M} statistically investigated the heights of the above-the-loop-top HXR and microwave sources. 
They found that the HXR sources tend to be located at higher altitudes than 17 GHz microwaves.
Figure \ref{fig:fig4} shows an example event, i.e., an X3.1 class flare occurring at the west limb on 2002 August 24 (detailed reports on this flare have been presented by \cite{2005ApJ...629L.137L} and \cite{2009ApJ...697..735R}).
It is reasonable to assume that electrons producing 17 GHz microwaves are mildly relativistic within a typical coronal magnetic field intensity of $\lsim 100$ Gauss \citep{1999spro.proc..211B}. 
Therefore, the observations imply that intermediate energy electrons (several tens to one hundred keV, seen as HXR sources) are located at a higher altitude than higher energy electrons (several hundred keV, seen as microwave sources) and lower energy electrons (up to several tens of keV, seen as coronal HXR loops).
Although the evaluation of emissions from the simulated electron distribution is necessary, the simulation result including pitch-angle diffusion is qualitatively consistent with observations of the energy-dependent height distribution. 
The result without diffusion cannot reproduce the observations because the electrons peak at the same altitude, regardless of their energy.
}}

{On the basis of the results, we consider that the above-the-loop-top HXR source is the region below which efficient loss takes place for electrons with the highest precipitation rate, determined by the balance of the bounce frequency in the loop and the pitch-angle scattering frequency.
Otherwise, these electrons continue to be trapped in the loop, follow the shrinking loop to lower altitudes, and peak at the same altitude observed in other energy bands.}


For numerical simplicity, we do not include the energy change due to Coulomb collisions in Equation (\ref{eq:2}). The collisional energy change is important for intermediate energy electrons, whereas it is not important for low energy electrons (with energy comparable to the temperature of target electrons) and for collisionless high energy electrons. Intermediate energy electrons may lose energy via collisions. From the viewpoint of particle loss and the resulting formation of the height distribution, the following holds in the case of the current simulation results: The loss takes place in either velocity or configuration space via deceleration or precipitation.

\begin{figure}[p]
\centering
\epsscale{.65}
\plotone{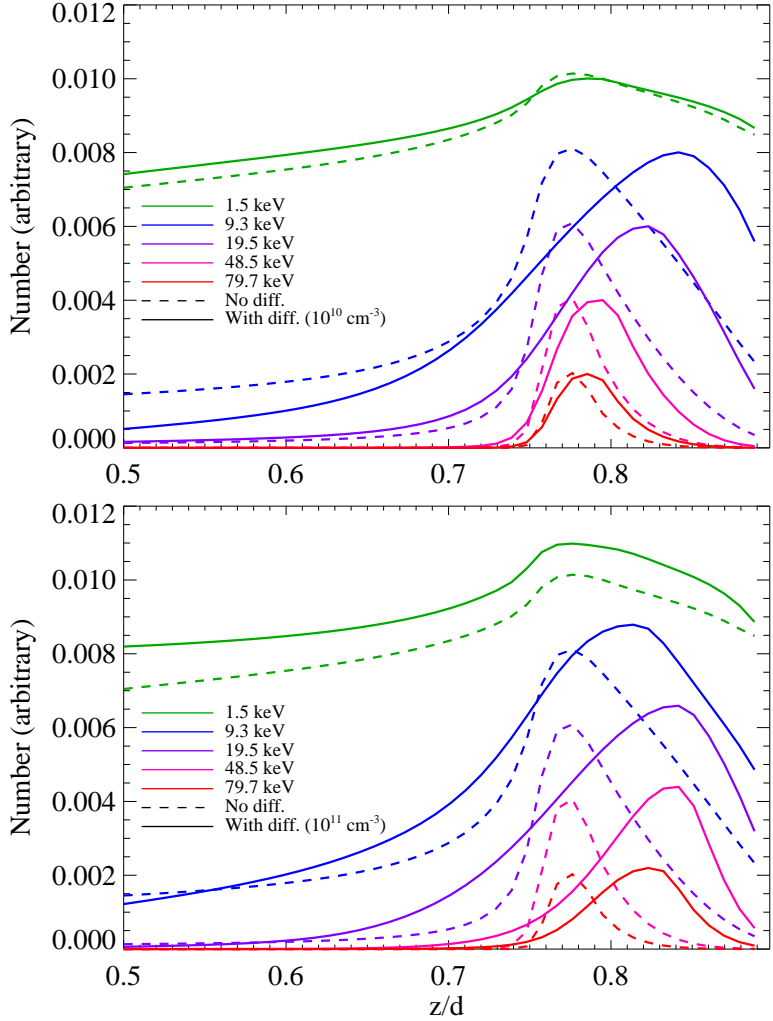}
\caption{Electron height distributions at the loop top ($x=0$). The solid and dashed lines represent the results with and without pitch-angle diffusion. The top and bottom plots show the results with initial number densities of $10^{10}$ and $10^{11} \; {\rm cm^{-3}}$, respectively. The green, blue, purple, pink, and red lines represent electron energies of 1.5, 9.3, 19.5, 48.5, and 79.7 keV, respectively. The electron number is multiplied by proper factors for illustration. The spatial coordinates are normalized by $d = 15 \; {\rm Mm}$.}
\label{fig:fig2}
\end{figure}

We have not achieved quantitative agreement between the observations and the model thus far. The above-the-loop-top HXR source is typically observed in the 20-50 keV energy band; on the other hand, with an initial number density of $10^{10} \; {\rm cm^{-3}}$, the current simulation has shown that the 10 keV electrons peak at the highest altitude and the 20-50 keV electrons peak at lower altitudes (Figure \ref{fig:fig2}, top). 
This is because betatron acceleration due to loop shrinkage significantly increases the number of high energy electrons toward a lower altitude, and overcomes the loss via precipitation into the footpoint. For further understanding the observations, we consider two possible scenarios.

\begin{figure}[p]
\centering
\epsscale{.65}
\plotone{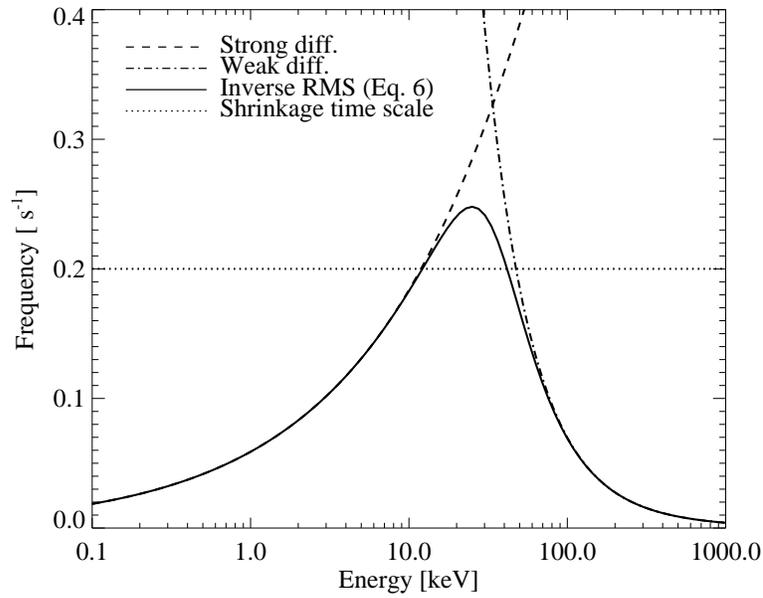}
\caption{Precipitation rate in the strong (dashed line) and weak (dash-dotted line) diffusion limits, and their inverse root mean square (solid line, Equation (\ref{eq:6})) as a function of the electron energy. {{A number density of $10^{10} \; {\rm cm^{-3}}$} is used.} The inverse of the time scale of loop shrinkage $(5 \; {\rm s})^{-1}$ is shown as a dotted line.}
\label{fig:fig3}
\end{figure}

\begin{figure}[p]
\centering
\epsscale{.65}
\plotone{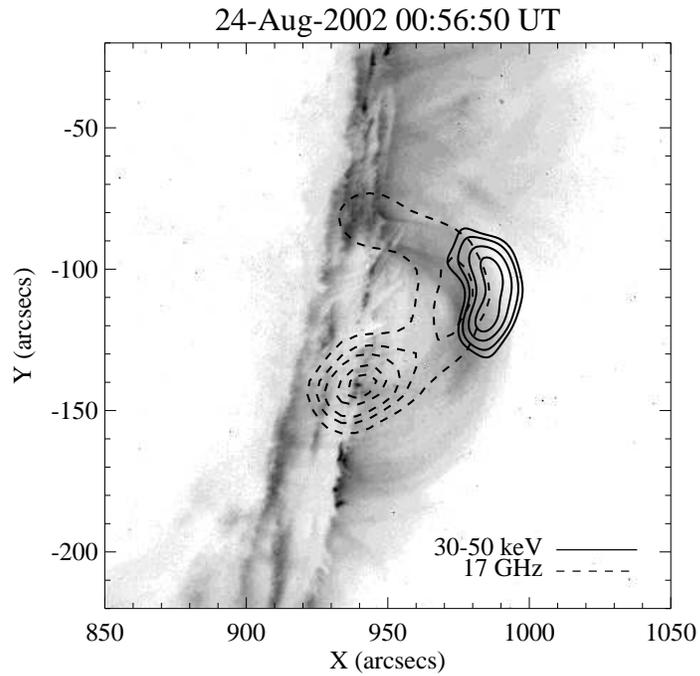}
\caption{Spatial distribution of {\RHESSI} 30-50 keV HXR (solid lines) and NoRH 17 GHz microwave (dashed lines) overlaid on the {\it Transition Region and Coronal Explorer} ({\TRACE}) 195 {\AA} negative image of the flare occurring on 2002 August 24. Contour levels are 50\%, 60\%, 75\%, and 90\% of the peak intensity for the HXR, and 20\%, 35\%, 50\%, 75\%, and 90\% for the microwave. }
\label{fig:fig4}
\end{figure}

In the first scenario, the rate of increase in the high energy electron number via acceleration is reduced if the slope of the energy distribution is flatter $(\partial \ln f / \partial t \propto -\partial \ln f / \partial \gamma)$. Thus, electrons should be primarily accelerated before the onset of shrinkage.
The initial temperature (3 keV) employed in the simulation seems to be insufficient to quantitatively explain the observations. If some mechanisms have primarily accelerated the electrons before shrinkage, the 20-50 keV electrons can potentially peak at higher altitudes than the 10 keV electrons because their precipitation rate is not sufficiently small to be approximated by the weak diffusion limit, even if it includes only Coulomb collisions (see Figure \ref{fig:fig3}). 
{{
Many possible scenarios have been proposed for primary acceleration, such as the reconnecting current sheet \citep{1996ApJ...462..997L} and the vicinity of the reconnection region \citep{2005JGRA..11010215H,2006Natur.443..553D}. 
Evidence of a current sheet has been observed, and hence, particle acceleration can potentially occur around this region \citep{2003ApJ...596L.251S,2004ApJ...612..546S,2008ApJ...676..704L}.
\cite{2008ApJ...676..704L} proposed a scenario wherein stronger (weaker) turbulence generated at a higher (lower) altitude may result in greater (less) acceleration and energy-dependent height distribution of coronal HXR sources. 
Although it seems reasonable to expect stronger turbulence at a higher altitude near the reconnection region, this scenario by itself is insufficient for producing microwave-emitting higher energy electrons at a lower altitude, where less acceleration is assumed.
We believe that a combination of primary acceleration and subsequent acceleration, transport, and dissipation processes rather than only a single acceleration process is crucial to the formation of the above-the-loop-top source.
 }}

In the second scenario, the loss of high energy electrons is enhanced if the precipitation rate is increased.
As seen in the bottom plot of Figure \ref{fig:fig2}, the 20-50 keV electrons are located at higher altitudes than the 10 keV electrons when the initial number density is $10^{11} \; {\rm cm^{-3}}$. However, we believe that this value is too high, and hence, unrealistic for a typical coronal circumstance. Alternative wave-particle interactions might take place and increase the precipitation rate, even if Coulomb collisions are inefficient.

{{
Let us roughly estimate the pitch-angle scattering rate via wave-particle interactions. 
High-frequency whistler waves are among the most probable agencies of electron scattering. The pitch-angle diffusion coefficient due to parallel-propagating whistler waves is approximated by
\begin{eqnarray}
D_{\mu \mu} \sim  \frac{\pi \omega_{ce} \left(1-\mu^2\right)}{2} \frac{k_{R} W(k_{R})}{B_0^2},\label{eq:1}
\end{eqnarray}
where $\omega_{ce}$ is the electron cyclotron frequency, $B_0$ is the background magnetic field intensity, $k_{R}$ is the resonance wavenumber, and $W(k)$ is the wave spectrum \citep{1966JGR....71....1K,1966PhFl....9.2377K}. The resonance wavenumber satisfies the following dispersion relation and cyclotron resonance condition,
\begin{eqnarray}
\left(\frac{c k}{\omega}\right)^2 = 1 - \frac{\omega_p^2}{\omega \left(\omega - \omega_{ce}\right)}, \;\;\; \omega - \mu v k = \frac{\omega_{ce}}{\gamma},\label{eq:4}
\end{eqnarray}
where $\omega_p$ is the plasma frequency and $c$ is the speed of light.
Assuming that it has a single power-law distribution, the wave spectrum is given by \citep{1992ApJ...398..350H,2000JGR...105.2625S}
\begin{eqnarray}
W(k) &=& \frac{q-1}{k_{\rm min}}\left(\frac{k}{k_{\rm min}}\right)^{-q} W_{\rm tot}, \;\;\; \\
 W_{\rm tot} &=& \inty{k_{\rm min}}{\infty} W(k)dk = R \frac{B_0^2}{8 \pi},\label{eq:5}
\end{eqnarray}
where $W_{\rm tot}$ is the energy density with a wavenumber greater than the cutoff wavenumber $k_{\rm min}$, and $R$ denotes the ratio of the wave energy to the background magnetic energy.
Assuming that whistler waves are excited via energy transfer from low-frequency {\Alfven} waves and that the transfer is much faster than cyclotron damping, we extrapolate the wave spectrum from the macroscopic scale to the whistler regime.
In the following, we use $k_{\rm min} = 2 \pi / 10^9 {\rm \; cm^{-1}}$ and $R=10^{-4}$.}}

\begin{figure}[p]
\centering
\epsscale{.65}
\plotone{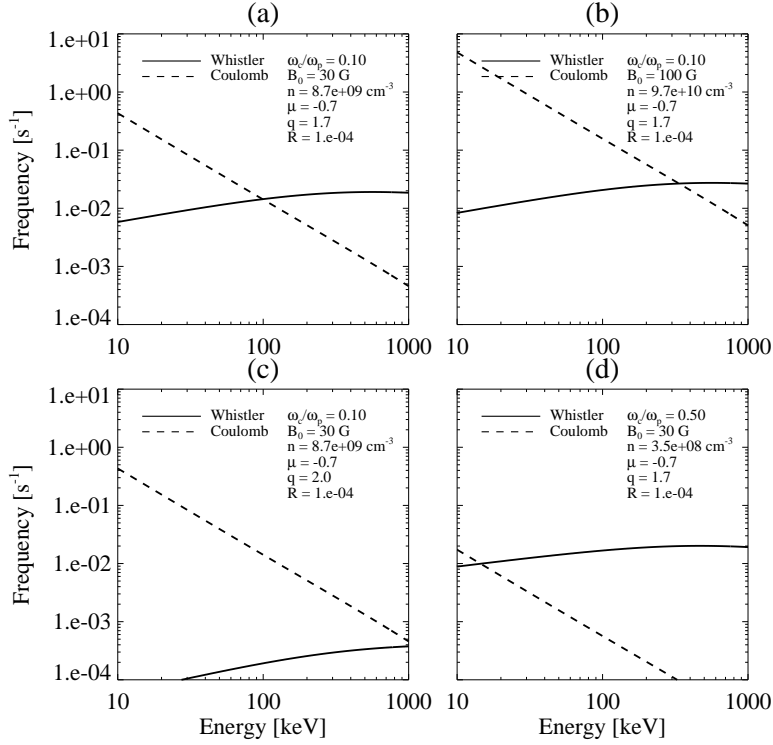}
\caption{Comparisons of pitch-angle diffusion coefficients due to Coulomb collisions (Equation (\ref{eq:3}), dashed lines) and whistler waves (Equation (\ref{eq:1}), solid lines) as a function of the electron energy. (a) Reference result with $(\omega_{ce}/\omega_p,B_0,n,\mu,q) = (0.1,30 \; {\rm Gauss},9 \times 10^{9} \; {\rm cm^{-3}},-0.7,1.7)$. (b)-(d) Results with higher magnetic field intensity and number density $((B_0,n) = (100 \; {\rm Gauss},10^{11} \; {\rm cm^{-3}}))$, steeper wave spectrum $(q=2)$, and higher $\omega_{ce}/\omega_p \; (0.5)$ with lower number density $(n=3.5 \times 10^{8} \; {\rm cm^{-3}})$, relative to the reference, respectively.}
\label{fig:fig5}
\end{figure}

{{In Figure \ref{fig:fig5}, we compare the pitch-angle diffusion coefficients due to Coulomb collisions (Equation (\ref{eq:3})) and whistler waves (Equation (\ref{eq:1})) on the basis of different parameters $(\omega_{ce}/\omega_p,B_0,n,\mu,q)$. Figure \ref{fig:fig5} (a) shows a reference result with $(\omega_{ce}/\omega_p,B_0,n,\mu,q) = (0.1,30 \; {\rm Gauss},9 \times 10^{9} \; {\rm cm^{-3}},-0.7,1.7)$. Higher energy ($\gsim 100$ keV) electrons are scattered by whistler waves rather than by Coulomb collisions; hence, they will be lost to a greater extent via precipitation than that expected in the current simulation. This may support the second scenario, which suggests that high energy electrons should be lost in the course of loop shrinkage. However, the pitch-angle diffusion coefficient due to whistler waves is intricately dependent on many plasma parameters. Figure \ref{fig:fig5} (b)-(d) shows results with higher magnetic field intensity and number density $((B_0,n) = (100 \; {\rm Gauss},10^{11} \; {\rm cm^{-3}}))$, steeper wave spectrum $(q=2)$, and higher $\omega_{ce}/\omega_p \; (0.5)$ with lower number density $(n=3.5 \times 10^{8} \; {\rm cm^{-3}})$, relative to the reference, respectively. 
The ratio of the pitch-angle diffusion coefficient due to whistler waves to that due to Coulomb collisions significantly varies according to the plasma parameters.
It is not easy to evaluate the contribution of wave-particle interactions to the total pitch-angle scattering rate in a realistic coronal circumstance.}}

{{
\cite{2010ApJ...714.1108K} reported a flare occurring on 2007 December 31, in which a coronal microwave source was cospatial with an above-the-loop-top HXR source, as opposed to the flare that occurred on 2002 August 24 (Figure \ref{fig:fig4}). On the basis of our model, their result may be interpreted as the occurrence of pitch-angle scattering and the resulting efficient precipitation into the footpoint at the same altitude {(just below the emission peaks)} for both intermediate and high energy electrons. It is plausible to consider wave-particle interactions rather than Coulomb collisions for efficient pitch-angle scattering over a wide range of energy (see Figure \ref{fig:fig5}) if sufficient waves are generated.
}}

The difference in peak heights with different energies is $\sim 1$ Mm in the simulation, whereas the observed above-the-loop-top HXR sources are located several Mm higher than the flare loops \citep{1995PASJ...47..677M,1996ApJ...470.1198A,1996ApJ...468..398A}. From a combination of loop shrinkage and the {\TPP} model (disregarding the effect of the acceleration), the difference is roughly evaluated as $v_{\rm loop}\left[ \tau - \left\{\max\left(\nu_{p}\right)\right\}^{-1} \right]$, where $v_{\rm loop}$ is the time-averaged shrinking velocity at the loop top, $\tau$ is the elapsed time of shrinkage, and $\max\left(\nu_{p}\right)$ is the maximum precipitation rate ($\sim 0.25$ in Figure \ref{fig:fig3}). If flaring electromagnetic fields are configured so as to sustain a fast flow $v_{\rm loop} \sim 1000 \; {\rm km \; s^{-1}}$ in $\tau \sim 10 \; {\rm s}$ (time scale of nonthermal emissions) {or if $\max\left(\nu_{p}\right)$ is considerably higher,} then the difference can be of the order of several Mm.
If such a condition is not realized, the difference will be $\lsim 1$ Mm, which is difficult to resolve spatially using current instruments. The rare detection of the above-the-loop-top source might indicate difficulty in sustaining the fast flow.


It is noteworthy to suggest that we can observationally determine the maximum precipitation rate if the above-the-loop-top HXR source and flow velocity are observed simultaneously. This is useful for discussing the detailed physical mechanisms of pitch-angle scattering, Coulomb collisions, and/or wave-particle interactions.
  
\begin{figure}[p]
\centering
\epsscale{.6}
\plotone{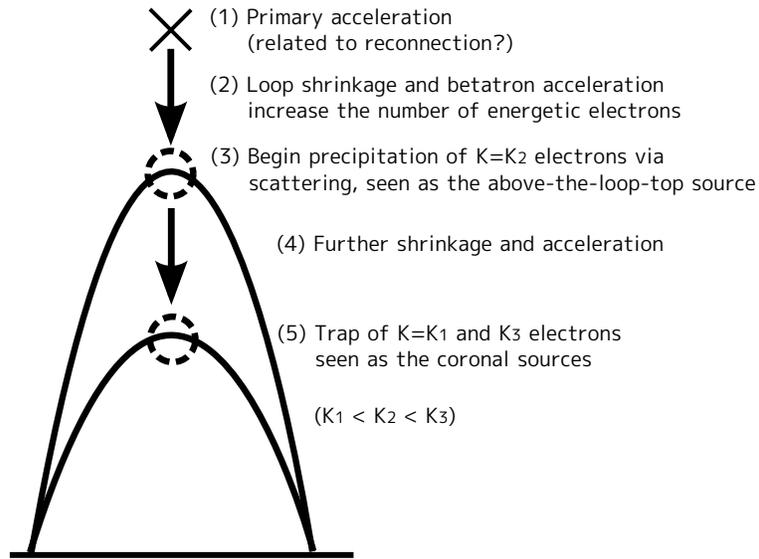}
\caption{Schematic illustration of the formation scenario of the above-the-loop-top HXR source. The solid curved lines denote magnetic field lines, the arrows denote the direction of the shrinkage of the loops, the dashed circles denote the emission sources, the cross denotes the magnetic reconnection site, and $K (K_1 < K_2 < K_3)$ indicates the electron kinetic energy. 
{By efficient precipitation via pitch-angle scattering, electrons with $K=K_2$ do not enter Step 4 and peak at a higher altitude in the course of loop shrinkage (seen as the above-the-loop-top source) than those with $K=K_1$ and $K_3$.}
 }
\label{fig:fig6}
\end{figure}

On the basis of these discussions, we suggest a formation scenario of the above-the-loop-top HXR source, as shown in Figure \ref{fig:fig6}. For the next step, we consider the acceleration near the magnetic reconnection region. 
Such a study can be combined with the drift-kinetic model, which will facilitate a better understanding of the particle acceleration in solar flares.

\begin{acknowledgements}
We would like to thank S. Imada for insightful comments on our manuscript, and an anonymous referee for carefully reviewing  the manuscript. This work was supported by a Grant-in-Aid for Young Scientists, (B) \#21740135; the joint research program of the Solar-Terrestrial Environment Laboratory, Nagoya University; and a Grant-in-Aid for Creative Scientific Research, ``The Basic Study of Space Weather Prediction'', from the Ministry of Education, Culture, Sports, Science and Technology (MEXT), Japan.
\end{acknowledgements}


\end{document}